\documentclass[runningheads]{llncs}
\usepackage{graphicx}
\usepackage{multirow}
\usepackage{threeparttable}
\usepackage{subcaption}
\usepackage{amsmath}
\usepackage[dvipsnames]{xcolor}
\usepackage{amssymb}

\newcommand{\name}{\textsf{Travelers}}

\newcommand{\routing}{routing}

\newcommand{\pompe}{\textmd{Pomp\={e}}}

\definecolor{ao}{rgb}{0.0, 0.5, 0.0}

\newcommand{\resolution}{$4k\Delta + 2\delta$}

\begin{document}
\title{{\name}: A scalable fair ordering BFT system}
%
%\titlerunning{Abbreviated paper title}
% If the paper title is too long for the running head, you can set
% an abbreviated paper title here
%
%\orcidID{0000-1111-2222-3333}
\author{Bowen Xue \and
Sreeram Kannan}
\authorrunning{B. Xue and S. Kannan}
% First names are abbreviated in the running head.
% If there are more than two authors, 'et al.' is used.
%
%\institute{}
\institute{University of Washington at Seattle
%\email{lncs@springer.com}\\
%\url{http://www.springer.com/gp/computer-science/lncs} \and
%ABC Institute, Rupert-Karls-University Heidelberg, Heidelberg, Germany\\
%\email{\{abc,lncs\}@uni-heidelberg.de}
}
\maketitle              % typeset the header of the contribution

\begin{abstract}
Many blockchain platform are subject to maximal value extraction (MEV), and users on the platform are losing money while sending transactions because the transaction order can be manipulated to extract value from them.
Consensus protocols have been augmented with different notion of fair ordering in order to counter the problem.
Out of all practical protocols, the most efficient BFT consensus requires $O(nTL + n^2T)$ communication complexity, where $n$ is number node, $T$ is number of transactions and $L$ is average transaction size.
In this work, we propose a new system of BFT fair ordering protocols, {\name}, that substantially reduce the communication complexity.
The proposed system of protocols satisfy a new notion of fair ordering, called probabilistic fair ordering, which is an extension to some existing notions of fairness.
The new notion allows a small probability of error $\epsilon$, that adversary can insert some transactions at any location in a block, but for the remaining $1-\epsilon$ the a modified version of ordering linearizability holds.
Our mechanism neither require a dissemination network nor direct submissions to all consensus nodes. 
The key innovation comes from a routing protocol, that is both flexible and efficient.
We construct a protocol with $O(c\log({n})TL + n^2)$ communication complexity with $\epsilon = 1/n^c$ for some system parameter $c\ge 1$.

\end{abstract}

\section{Introduction}
\label{sec:introduction}

Cryptocurrency transaction has been a major application to blockchain for more than ten years since the start of Bitcoin.
With the programmable features, Ethereum has enabled an array of decentralized financial (Defi) applications including exchanges, lending platform and stable coins. 
A great amount values are transacted every day, the daily trading volume on Ethereum is more than five Billions US dollar\cite{ethVolume}.
With so much values attracted into the blockchain, people starts to exploit some unspecified area in the protocol to extract Maximal Extractable Value (MEV) from users.
The extraction is possible because most consensus protocols used today does not concern about the ordering of transactions. 

For example, automated market maker (AMM) is an type of programmable exchange on blockchain.
If a user sends a transaction to purchases asset,
adversary can make profits out of user by placing two transactions before and after the user' transaction.
Such attack is called sandwich attack, and has been extensively documented and analyzed in the past, see \cite{daian2019flash,Heimbach2022}.
Besides the sandwich attack, people have discovered various other attacks specific to the Defi protocols, see \cite{yang2022sok} for a more comprehensive list.

To counter the problem, people have proposed at least three categories of solutions.
First, notions of fair ordering have been proposed by many consensus protocols, including \cite{cachin2022quick,kelkar2021themis,zhang2020byzantinepompe,constantinescu2023fair,lyra}. 
Most of them align on a common FIFO principle: a transaction perceived by the protocol first, should be executed before others, though the way they measure the ordering are different.
%
%new property beyond the conventional properties of Byzantine Fault Tolerant(BFT)\cite{cryptoeprint:2023/397} protocols. This captures a sense of fairness about the order of transactions.
%
%The property is important, including automated market maker (AMM), onchain Dutch Auction and liquidation. 
%
%Fair ordering is important for many blockchain application, because it can inhibit bad actors from front-running any transactions. It is especially important for transactions of applications that are sensitive to locations in a blockchain. 
%
Second, many large projects including Ethereum and application-chains from Cosmos adopt an approach called, MEV auction platform. 
With this approach, MEV is not alleviated, but optimized to extract the greatest value and then redistribute the revenue back to the block proposers who are parts of ecosystem.
The third approach can be categorized as content-agnostic ordering\cite{yang2022sok,ope}, essentially the transaction content is committed and threshold encrypted to hide it from the adversary, so that by the time it is revealed, the transaction is already confirmed on the blockchain. 
Compared to fair ordering protocol, the content-agnostic approach offers a weaker guarantee, because it provide no ordering properties about its transactions. 
For example, a transaction submitted much later in a block can still front-run others transaction.
%
%Paper\cite{yang2022sok} provides a good summary about current status on the MEV countermeasure.

%In Etheureum, a solutoin called MEV-boost allows a single block proposer to unilaterally decide the the order of transactions in a block.
%
%The economical incentive for those proposer is to extract \textbf{MEV} (maximal extractable value) arbitrarily reordering the transactions to maximize its own utility.
%
%Indeed, there is a flourishing ecosystem built on top of foundation of economic incentives through auction.
%For example, Flashbot 
%
%In the future, Ethereum has considered a roadmap called Proposer-Builder Separation to incorporate MEV inside the protocol.
%

In the current blockchain ecosystem, the MEV auction approach has demonstrated to be the most popular. 
There is a flourishing ecosystem built on top of foundation of economic incentives, that includes Flashbot\cite{flashbot}, Bloxroute\cite{bloxroute}, Blocknative\cite{blocknative}, Eden\cite{eden}, Skip Protocol\cite{skip}.
%
%that include large projects like Ethereum and many application chains inside Cosmos ecosystem.
%
%Without regarding the morale and implication for MEV auction,
The auction solution has a great advantage of $O(1)$ communication complexities. 
This is achievable because only a few centralized entities need to collect transactions in order to build a best block to win the auction. 
The second popular approach is the content-agnostic ordering. Although it offers less properties than fair ordering protocols, many projects like Shutter network\cite{shutternetwork}, Sikka\cite{sikka} have implemented and started to contribute the blockchain ecosystem, because of its simplicity and wide spectrum of applications.
With regard to the category of fair ordering protocols, till now as far as author is aware, there is no well known project that operates a fair ordering protocol (except Chainlink has promised it since September 2020\cite{chainlinkfss}). 

We notice all existing fair ordering protocols have rather stiff ordering policy enforcements regardless of the underlying fairness notions.
Out of all practical solution, Themis\cite{kelkar2021themis} is the most communication efficient protocol, but still requires $O(nTL + n^2T)$ communication complexity, where $L$ is the transaction size, $T$ is number of transactions, and $n$ is total number of nodes\footnote{Themis SNARK\cite{kelkar2021themis} has $O(nTL + nT)$ communication complexity. But SNARK requires cryptographic computation order of magnitude higher than commitments or hashes. It serves only the theoretical purposes, as self-described by \cite{kelkar2021themis}}.
Importantly, Themis consensus protocol is entangled with complexity to deal with Condorcet Cycle, which is a phenomenon that the ordering among transactions are chained together to form nested loops, as the result those transactions have to be batched together. 
Although the protocol is carefully designed, the resulting protocol entails a great complexity.

This work explores a new notion of fair ordering,
referred as probabilistic fair ordering.
The purpose is to relax the criteria of existing fair order, by removing the requirement to have all nodes receiving identical transactions.
%
%For a fair ordering notion, the new notion ensures the properties relating to the notion hold with probability $1 - \epsilon$, for some small $\epsilon$.
%
We then introduce a class of protocols, called {\name}, based on a new notion called probabilistic ordering linearizability.
Ordering linearizability is proposed by {\pompe}\cite{zhang2020byzantinepompe},
that requires a real clock on each node to lock a transaction with a timestamp.
Because timestamps are linear, all locked transactions with timestamps can be totally ordered by a simle sorting. It is a key insight mentioned in {\pompe}\cite{zhang2020byzantinepompe} that avoids the complex puzzle of Condorcet cycle.
But even with a clock, {\pompe} still requires a communication complexity of $O(n^2TL + n)$.
With the new relaxed notion of probabilistic ordering linearizability, {\name} can achieve $O(c\log({n})TL + D)$ communication complexity with error probability $\epsilon=\frac{1}{n^c}$ for some system parameter $c$, where $D$ is the total communication complexity for some consensus protocol.
In {\name}, the precision of fairness notion depends on how large the mismatch $\delta$ among nodes' clock and network latency $\Delta$. 
We elaborate on the probabilistic notion in Section~\ref{sec:fair_ordering} and provide a concrete fairness definition for one {\name} protocol in Section~\ref{sec:travelers:properties}.

\subsection{{\name} system}
\label{sec:introduction:travelers}

At the high level, we can use a framework to split the system into two components: transaction submission and consensus. 
Transaction submission is the part responsible for getting sufficient nodes to receive the transaction, and it is where the transaction size $L$ shows up in the complexity analysis. 
The consensus part is responsible for making agreement on some data critical to the consensus protocol. 
We will further elaborate this framework in the next Section.
Unlike other fair ordering protocol, the submission component in {\name} neither requires a P2P dissemination network, nor it is required to send transactions to all correct nodes. 
Reducing the dissemination phase is important to improve the first term of the communication complexity from $O(nL)$ to sublinear in $L$, see Table~\ref{table:ordering-notion}. 

\begin{figure}[h]
\centering 
\includegraphics[width=\textwidth]{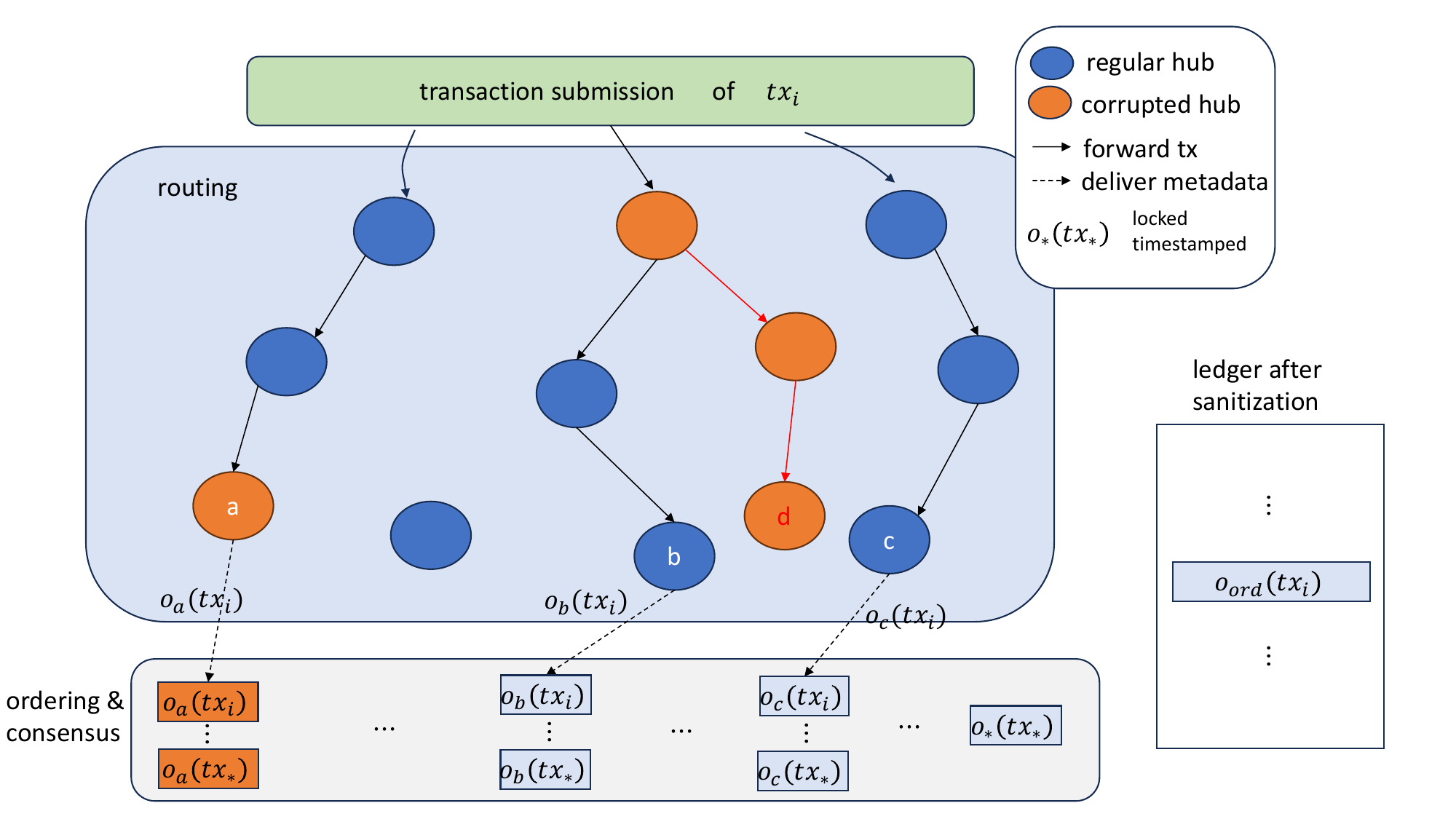}
\caption{Travelers architecture. Suppose every hub consists of a single node. Each of three copies of transaction has distinct path. Nodes $a,b,c$ are the nodes from the final hubs. The path containing node $c$ consists of regular hub only. We use $o_{ord}$ to denote the locked timestamp.}
\label{fig:traveler_architecture}
\end{figure}

In our submission component, we design a routing protocol that moves a transaction through a series of hubs on a path. %, where each has only a few nodes.
The transaction receives timestamp at each hub, and all timestamps are used together by the {\routing} protocol to produces the locked timestamp, which is used for ordering.
Figure~\ref{fig:traveler_architecture} displays several paths each contains three hubs.
%After a client sends copies of the identical transaction to three random node hubs, the {\routing} protocol 
%
A hub is a collections of nodes, whose membership and size are determined by some assignment function decided by the protocol. 
%
%The assignment function is a system choice, and the possible size for a hub can be any number from one to all $n$ nodes. 
%
We will discuss more about the assignment function from Appendix~\ref{appendix:route:assignment}.
However, in Figure~\ref{fig:traveler_architecture} each hub is assigned with only one unique node.

When a transaction arrives, a hub timestamps its arrival time and approves it with a cryptographic signature on both metadata and the transaction payload.
A corrupted hub indicates the set has significant amount of malicious nodes, such that they alone can approve the transaction with arbitrary timestamp. On the other hand, if honest nodes alone can approve the transaction, it is called the regular hub.
%We will discuss more in Section~\ref{sec:route:hubs_timestamps_type}.
%
%In each hub, signatures from sufficient members from the hub is needed for hub's approval.
%
If all of hubs in a path approved the transactions, 
the nodes from the last hub deliver all timestamps to the consensus along with signatures and transaction payloads.
We define maximal timestamps in the path as the \textbf{locked} timestamp for the transaction on that path.

In Figure~\ref{fig:traveler_architecture}, we show a red path consisted entirely of malicious hubs.
It is referred as the corrupted path, and similarly a path of all regular hub as the regular path.
We want to decrease its probability of corrupted path such that it is less than $\epsilon$.
Because if no regular hub can bound the timestamp, an adversary can lock arbitrary timestamp for any transactions that travel the path.
Not only we need to lower $\epsilon$, the routing protocol also needs to cap the total possible number of paths by using a deterministic random function.
Because there are combinatorial number ways to arrange hubs into distinct paths, no matter how small $\epsilon$ be, an adversary can find a corrupted path if given sufficient chances.

When a client sends multiple copies of the same transaction to traverse distinct paths, each would generate a different locked timestamp.
It is the ordering rule's responsibility to determine the canonical timestamp for the transaction.
%
%In contrast, timestamps generated from paths of any corrupted hubs can be tampered by malicious node to either delay or speedup delivery.
%
%As the result, the routing protocol needs to be designed to ensure there exists an honest path with high probability.
%
%In Figure~\ref{fig:traveler_architecture}, all certified timestamps along with transactions are stored locally by each node. 
%
Once one canonical timestamp is selected by the ordering rule, because timestamps are linear by nature, the total order of transactions can be derived by a simple sorting algorithm.
The right side of Figure~\ref{fig:traveler_architecture} provides global view about the ledger after the sanitization.
%
%But instead of running a standard BFT consensus that replicates everything on each node,
%
%we can run a much communication efficient version with erasure code.
%

The notion of ordering linearizability decouples the ordering and consensus, as pointed by many works\cite{zhang2020byzantinepompe,Heimbach2022,lyra}. 
Because timestamps lock transactions, the only part we need from the standard consensus is just an agreement on a stream of bit, which is data availability, so that 
anyone can be retrieved them later.
%
%Everyone can download the data and derive the canonical ordering locally, and this removes great complexities that modify the consensus protocol to handle Condorcet cycle.
%
%Essentially, the only needed property is the data availability that allows a group of nodes to reach consensus on a stream of bit and retrieve later.
%
Standard asynchronous and partial asynchronous BFT protocol inherently satisfy data availability, but make it a requirement for every node to download the entire data during the agreement protocol. 
%
%To reduce the complexity further, Erasure code can be used to remove this requirement, and therefore lower the communication complexity further.
%
For fair ordering protocol, \textbf{censorship resistance} is an important consensus property but missed by many standard leader based protocol .
It guarantees some transaction is included in the next block even if the current and future leaders have the intent of censoring.
{\name} requires the underlying consensus to have this property in order to commit the locked timestmap.
Most standard leaderless BFT protocol has censorship resistance in nature.
For example, DispersedLedger\cite{yang2022dispersedledger} is an asynchronous protocol that uses Erasure code to improve throughput.
VABA\cite{abraham2018validated}, Dumbo-MVBA\cite{mvba} are leaderless protocols that have communication complexity of $O(n^2)$. 
Because timestamps lock transaction before consensus, the confirmation latency required by the consensus protocol does not affect final ordering.
%but suffers a $O(n^2)$ communication complexity due to all-to-all communication pattern.
%
%But it is possible to lower the complexity further with a leader based partial asynchronous protocol.
%In order to improve the complexity further,
%
%we provide an outline on how to modify Hotstuff-2\cite{hotstuff2} with small changes to incorporate both censorship resistance and Data availability with $O(n)$ complexity.
%
%incorporate data from $O(n)$ nodes and provide only data availability.
%
%The final consensus part requires $O(n\lambda)$ with a property of censorship resistance. 
\textbf{BigDipper}\cite{xue2023bigdipper}, contains more rigorous formulation and presentation, and shows that the complexity can be further reduced to $O(\log(n)\lambda)$, resulting to a protocol with only $O(c\log(n)TL + \log(n)n\lambda)$ overall communication complexity.

Similar to many protocols based on the notion of ordering linearizability, {\name} can further protect transaction with encryption.
The idea is to let the client encrypt the transaction, so that the ciphertext can only be decrypted toward the end of the path, such that no adversary has sufficient time to frontrun the transaction. We discuss it in Section~\ref{sec:route:encrypt}.

\subsection{Comparison with other approaches}
\label{sec:introduction:fair_order_comparison}

%The MEV problem has attracted interests from both industry and academia. 
%
%Currently the most popular MEV solution is the MEV auction platform. Many large projects including Ethereum and application chains from Cosmos ecosystem adopt this approach. 
%
%Using the approach, MEV is not alleviated, but optimized to redistribute the reward to the block proposer who is a member of the nodes. 
%

%
%Content-agnostic ordering approach uses a commit-and-reveal approach such that the transaction is encrypted before confirmation, and then is decrypted only afterward.

%Given no project has implemented fair ordering protocols, compared to other approaches, 
%
%this paper provides a MEV countermeasure that is weaker than the traditional protocols by making it probabilistic.
%

There are two general approaches inside the fair-ordering category. 
Almost all of them can be split into the two components we used in the last section.
The first approach uses the notion of ordering linearizability, represented by protocols including {\pompe}\cite{zhang2020byzantinepompe}, Decentralized Clock Network(DCN)\cite{Heimbach2022} and Lyra\cite{lyra}.
At the high level, every transaction first goes through a 
transaction submission component in order to lock a timestamp.
%protocol in order to achieve Byzantine ordered consensus (BOC), which produces a timestamp for each transaction.
%
For every transaction, {\pompe} devises its broadcast protocol to ensure everyone agrees on the median timestamp with $O(n^2)$ complexity.
DCN created a timestamp agreement protocol that has $O(n^3)$ complexity.
Lyra designs a Validated Value broadcast protocol based on Binary Agreement protocol\cite{mostefaoui2014signature} which take $O(n^2)$ complexity.
All of them requires all nodes to receive the transaction.
For the consensus component, they use some modified version of standard consensus protocol for replicating transaction along with the timestamp. 
Both Lyra\cite{lyra} and DCN\cite{Heimbach2022} claim they do not assume a synchronous clock.
Those protocol can easily incorporate threshold encryption 
like content-agnostic approach to provide further front-running protection.

The second common notion for fair ordering is batch order and differential order fairness \cite{cachin2022quick,kelkar2022order,kelkar2021themis}. They rely on relative transaction ordering to derive the final canonical ordering.
In the transaction submission component, all transaction is assumed to be propagated in a dissemination network which ensures all honest nodes to observe everything eventually.
This is the necessary requirement for those protocols, so that every honest node can create its local logical ordering to decide what can be ordered safely.
In the consensus component, because there are possible Condorcet cycle among the transaction, the consensus protocol needs to handle more works to ensure only subset of observed transactions can be confirmed.

%A recent paper\cite{ope} classify those as data-independent protocols, which only depends on the metadata of the transaction. 
%
%In {\pompe}, the metadata is encoded as the medium timestamp; in Aquetas, Themis or quick order fairness, the metadata is the relative logic order among other transaction. 
%

%
%Unlike {\pompe}\cite{zhang2020byzantinepompe} and DCN\cite{Heimbach2022} that uses median timestamp, we choose to use the earliest timestamp.
%
%In section~\ref{sec:route}, we delineate the reasoning and tradeoff behind it.
%
%We use $o_*(tx_*)$ to denote the final certified timestamp.
%In all existing fair ordering protocol\cite{cachin2022quick,kelkar2022order,kelkar2021themis,zhang2020byzantinepompe,lyra,Heimbach2022} mentioned above, the transaction submission component either assume a 
%
%the {\routing} protocol is constructed by a broadcast network that is either realized by a P2P network or direct unicast to all BFT nodes.
%
%Flooding transaction to everyone is a fundamental requirement for those fair ordering module so that each node can derive a local ordering to confirm with the canonical final ordering.
%

%However for each transaction, this requires all $n$ BFT replicas to receive identical data. It can be implemented as $n$ unicast connections or through a dissemination peer-to-peer network which requires higher effort of maintenance.
%
%Ideally we want a BFT protocol that requires only a few replicas to receive the transaction while still providing some ordering fairness.
%
%The key bottleneck of existing solution resides the nature  of everyone receiving everything.
%

Table~\ref{table:ordering-notion} summarizes a comparison among each algorithm, and its analysis is presented in
Appendix~\ref{appendix:complexity}. 
Like Quick Fair Ordering\cite{cachin2022quick}, we include the transaction size $L$ in the table, since transactions might take significant size as they become more complex. The number of transactions $T$ is a part of analysis, because clients' traffic is out of protocol control, this is beneficial to take it into consideration to understand how system behave in response to workload.
$\lambda$ denote the size of signature and hash value, and is typically $32$ bytes.
Compared to others, {\name} reduces communication complexity in both submission and consensus component.
{\name} is the first fair ordering protocol that achieves sublinear communication complexity in the transaction submission phase, it is because we use probabilistic fair ordering that allows some $\epsilon$ error.
Since {\name} is family of protocols, in Table~\ref{table:ordering-notion} we include two versions of it. 
Travelers-Speed is optimized for faster latency to traverse a path. 
Travelers-Light is optimized for the lowest communication complexity.
%
%With Erasure code, the consensus part of {\name} is reduced to $O(n\lambda)$, which is also significantly better than others. 
%
Not only {\name} is efficient, it is also flexible because both the routing protocol and the BFT consensus can be replaced easily by protocols with similar properties.
For example, all DAG based protocols can use Travelers to get fair ordering, like BullShark\cite{spiegelman2022bullshark}, Tusk\cite{danezis2022narwhal}.
In the routing protocol, both hub size and path length can be tuned; the assignment functions can also be designed to work with a topology consisntent to distributed validators technology like Obol\cite{obol} to achieve decentralization and threshold encryption.
%
%The consensus component can be replaced as long as they satisfy agreement on a stream of bits and censorship resistance.

%
%{\name} is designed to modular in order to maximally expose interface for better interoperability. The main part of the paper delineates the components, we write out the full protocol in the appendix.

%{\name} is a BFT system that supports fair ordering properties
%based on a notion of order linearizability.
%
%Like {\pompe}, our order linearizability requires each transaction be associated with a timestamp, such that all transactions can be sorted in the most straightforward sense. 
%

%It requires us to assume a synchronized clock so the timestamps are comparable to each other. 
%
%The assumption is different from the network assumption which specifies how long a message can be eventually delivered. 
%

%It seems very similar to the committee design pattern, which has been explored before by using sampled smaller committee like omniledger. But we use it for 

%
%A transaction must go through a series of hub in order to receive a final certificate that permits itself into the consensus protocol.

%The table contains a high level abstraction about the required assumption for each protocol. Each protocol contains finer details. Themis paper indicates that {\pompe} suffers a censorship attack, and quick-ordering fairness does not have liveness unless all nodes are honest. 

\begin{table}[!hbt]
\caption{Difference of ordering notion and complexity}
\begin{subtable}[h]{\linewidth}
\resizebox{\textwidth}{!}{
\begin{tabular}{|c | c | c  | c | c | }
\hline
\multirow{3}{*}{} Fairness & Algorithm & Security & Tx Submission & Total  \\
 Notion & Complexity & Assumption & Per payload & $T$ Payload  \\
 & & & Complexity & Complexity \\
 \hline\hline
 Batch-Order-Fairness\cite{kelkar2021themis} & Themis & $1/4$ & $O(nL)$ & $O(nTL + n^2T \lambda)$  \\
 \hline
 Differential-Order-Fairness\cite{cachin2022quick} & Quick o.-f. & $1/3$ & $O(n^2L + n^3\lambda)$ & $O(n^2TL + n^3T\lambda)$\\
 \hline
 Ordering Linearizability\cite{zhang2020byzantinepompe} & {\pompe} & $1/3$ & $O(nL + n^2\lambda)$ & $O(nTL + n^2T\lambda +  n\lambda)$ \\
 %\hline
 %Prob. Ordering Linearizability & Travelers-Single & $1/3$ & $O(c\log(n)n^{0.369(c+1)}L)$ & $O(c\log(n)n^{0.369(c+1)}L + n\lambda$)  & \greencheck \\ 
 \hline
 Prob. Ordering Linearizability & Travelers-Speed & $1/3$ & $O(c\log^2(n)L)$ & $O(c\log^2(n)TL + n^2 \lambda$) \\ 
 \hline
 Prob. Ordering Linearizability & Travelers-Light & $1/3$ & $O(c\log(n)\lambda+L)$ & $O((c\log(n)\lambda+L)T + \log(n)n\lambda)$)\footnote{By using iterative routing from Appendix~\ref{appendix:route:non_singular:traversal}, and 
 BigDipper for consensus} \\ 
 \hline
\end{tabular}
}
\end{subtable}
\label{table:ordering-notion}
\end{table}

%We propose a novel relaxed notion of fair ordering that, for every block there is some probability $\epsilon$ that fair ordering is violated, but the remaining $1-\epsilon$ probability the system maintain a more traditional fair ordering such that if a transaction $tx_a$ is sent before another transaction $tx_b$ at least for a duration of $h \Delta_r + \Delta_s$ , $tx_a$ is ordered before $tx_b$, where $h$ is one of the system parameters that determines the probability of $\epsilon$.

%Suppose for a transaction $t^a$, and trace generated for it from node $i$ is denoted as $tr^{a}_{i}$. Existing protocols requires a set of $tr^{*}_{j}$ for sufficient nodes $j \in J$ where $J \subseteq \{1..n\}$. 
%
%However, it is inefficient to flood the network. We first relax the traditional notion of fair ordering and derive {\textbf{Probablistic Fair Ordering}}

%Themis\cite{kelkar2021themis} and Aequitas\cite{kelkar2022order} use the notion of batch order fairness to derive the order of a canonical chain.
%
%Pompe\cite{zhang2020byzantinepompe} uses the notion of order linearizability, and requires synchronized clock (but it is different from the assumption that all messages are delivered in bounded time).
%
%Quick Fair ordering\cite{cachin2022quick} uses the notions of differential order fairness. Quick Fair ordering\cite{cachin2022quick} paper presents a nice table.

In Section~\ref{sec:route:hubs_timestamps_type}, we delineates the adversary action spaces, and in Section~\ref{sec:travelers:strategy} we reason about how {\name} defends against MEV attacks.

\section{Background}
\label{sec:background}

\subsection{BFT problems}
\label{sec:background:bft}

A Byzantine Fault Tolerant(BFT) protocol is a system of $n$ nodes that reaches consensus on a common state.
Clients submit transactions to some honest nodes in order to modify the state.
To reach agreement, nodes communicate among in a network which can be modeled as asynchronous, partial asynchronous and synchronous.
Most real system choose to use either asynchronous or partial asynchronous network model.
In a partial asynchronous network, there is a notion of Glocal Stabilization Time, after which all transaction arrive at known bounded duration.
Whereas in the asynchronous network model, messages among nodes are eventually delivered, but its time is unknown.
A BFT protocol has to satisfy the safety and liveness properties:
\begin{itemize}
    \item Safety: At any time, the ledger of every pair of honest nodes is a prefix of another
    \item Liveness: Transactions submitted by honest clients are eventually added into the ledger of honest nodes.
\end{itemize}
There is a lower bound for the number of malicious nodes. For most standard BFT protocol, it is $n \ge 3f+1$. For Themis, it is effectively $n \ge 4f+1$.

\subsection{Condorcet Cycle and Relaxation}
\label{sec:background:cycle}
Drawn from Social choice theory, a Condorcet cycle is a phenomenon that groups of transitive relation results into a final non-transitive relation. For example, suppose three voters $a,b,c$ rank three commands, $c_1, c_2, c_3$. A Condorcet cycle arise if the following ranks are given: 
$c_{1}^{a} \prec c_{2}^{a} \prec c_{3}^{a}, c_{2}^{b} \prec c_{3}^{b} \prec c_{1}^{b},  c_{3}^{c} \prec c_{1}^{c} \prec c_{2}^{c}$, where the superscript denotes the voter and $\prec$ is some transitive relation.
At the end, there is no conclusion which command is ranked the highest.  
Due to this fact, previous work\cite{Aequitas} shows that it is impossible to define fairness based on reception order and instead define batch-order-fairness, such that all transactions in the cycle are delivered altogehter.
However, sometimes when there is Condorcet cycle spanning across multiple blocks intervals, a transaction might have been already recognized by the consensus, but needs to wait for multiple blocks before confirming the transaction.
%Condorcet cycle which would invalidate their notion of batch-order-fairness.
%
It is due this fact, the consensus protocol based on batch order fairness needs not only just the data availability, it needs to decide what transactions are allowed to confirm.
Ordering linearizability avoids this problem and has been pointed by {\pompe}, Lyra, DCN\cite{Heimbach2022,lyra,zhang2020byzantinepompe}.
In Themis, nodes and the leader exchange not only the transactions list, but also the leader is responsible for aggregating the updates messages from the nodes.
In order to prevent a violation to a weak notion of liveness,
%So in batch-order-fairness, transactions cannot be confirmed immediately even after received by $f+1$ nodes, because there are cases that cycle are not fully resolved, and needs to wait after next block to wait for more nodes to receive the transaction. 
special techniques like unspooling or deferred ordering are introduced to Themis.

\section{Probabilistic Order Fairness}
\label{sec:fair_ordering}

We introduce a high level notion called Probabilistic Order Fairness.
%
%At the core of the concept, it adds a probabilistic attribute to some fair ordering notions. 
We start by introducing Ordering Linearizability, and based on that we state where the probabilistic component arise in the new notion.
%
%For instance, suppose for \textbf{Probabilistic Ordering linearizability}, informally it means there is $\epsilon$ chance that all transactions inside the BFT block failed to adhere to the notion of ordering linearizability. But for the rest of $1-\epsilon$, all transactions are sorted correctly. 
%
%Because we are attaching a statement about a certain protocol, the notion of probabilistic protocol is more general. We can think of {\pompe} as an protocol that implements the ordering linearizability with $\epsilon=0$.
%
%In the most general sense, \textbf{Probabilistic Fair Ordering} can be understood as: 
%
%\begin{itemize}
    %\item For every new block in the BFT protocol, there is $\epsilon$ probability that the stated notion is not achieved for the block, but in the remaining $1-\epsilon$ probability, the desired properties derived from the notion holds.
%\end{itemize}
%
%There are some differences about two uncertainties. Uncertainty on inclusion of one transaction has a local effect on the one transaction. But the uncertainty in probabilistic order fairness can affect all transactions in the block. 
%
%More concrete definition can be derived by applying the probabilistic statement to a specific notion. 
%
%Ordering Linearizability is proposed in the paper of {\pompe}\cite{zhang2020byzantinepompe}.
%
In simple term, Ordering Linearizability requires that 
\begin{itemize}
    \item if the highest timestamp of a transaction $tx_a$ provided by any honest nodes is lower than the lowest timestamp of a transaction $tx_b$ provided by any honest nodes, and if both transactions are committed, then $tx_a$ will occur before $tx_b$ in the final totally ordered ledger.
\end{itemize}

In {\pompe}, it is implemented by locking timestamps for all commands, so that the ordering linearizability can be enforced.
To lock a transaction, all nodes need to first receive the transaction, then broadcasts and agree on the median timestamp, which is both upper and lower bounded by timestamp generated by honest nodes.

%and undergone a procedures called {\textbf{Ordering Phase}}. The phase has a purpose to 
%
%When choosing what timestamp to lock for a transaction, the median timestamp is specifically chosen for its nice property that the timestamp must have been generated by an honest node.
%
%Since every node receives the transaction and given standard BFT assumption that has $2f+1$ honest nodes. 
%
%It is always the case that at least $2f+1$ timestamps are generated and collected by the protocol of {\textbf{Ordering phase}}.
%
%Out of collected timestamps, although $f$ timestamp can be arbitrarily manipulated by adversary, it can be shown that the median timestamp is always both upper and lower bounded by timestamp generated by honest nodes. 
%
%At least, all transactions are sorted based on the locked median timestamps. 
%
However, when {\pompe} defines the fairness, it is implicit that every node has a vote on the final results, which is achieved by having everyone receiving the transaction.
In our case, only nodes along the path will receive the transaction,
and there is a case that some transaction whose delivery is entirely determined by the malicious hubs. 
So in the probabilistic version, we made the following modification based on the previous definition:
\begin{itemize}
    \item For every new block in the BFT protocol, there is $\epsilon$ probability that some transaction $tx_c$ provided by a corrupted path can be added into the block at any locations, for the remaining $1-\epsilon$ probability, if timestamp of a transaction $tx_a$ provided by a regular path is $x$ duration before the timestamp of a transaction $tx_b$ provided a regular hub, and both transactions are committed, then $tx_a$ will occur before $tx_b$ in the final totally ordered ledger.
\end{itemize}
where $x$ is the duration parameter affected by the length of the path, clock mismatch among all nodes and exact routing algorithm. We elaborate in Section~\ref{sec:route}.

{\name} is a set of possible realization for such probabilistic notions, and it is likely other systems might achieve the similar properties. 
%
%In {\name}, the probabilistic element comes from the accuracy of timestamp. 
%
%In {\pompe}, the timestamp is always accurate in the sense it is both upper and lower bounded by some timestamps from honest nodes.
%
Compared to {\pompe}, {\name} relaxes the assumption that everything is received by everyone, hence it has to add new definition to cover the case. 

\subsection{Error Interpretation and Probability in Consensus}
The probabilistic notion relax the condition which a protocol has to satisfy, and therefore allows us to simplify protocol.
The error probability $\epsilon$ can be set in two ways depending on the interpretation.
Algorithmically, we can set the target $\epsilon$ to be negligible by letting $\epsilon=O(1/n^c)$ where $n$ is the number of nodes. 
We can also interpret $\epsilon$ economically.
Suppose the cost of attack per block is $C$, and the revenue from block reordering is $D$, then as long as 
$\frac{C}{\epsilon} >> D$, there is no benefits for adversary to initiate an attack.

We note that there has a history to introduce probability into the consensus properties.
In the classic partial synchronous BFT protocols, like Hotstuff, PBFT, the liveness property guarantees an eventual inclusion of a transaction depending on if the next leader is honest, which is a probabilistic statement.
For asynchronous BFT protocols, both safety and liveness are probabilistic requiring a random coin for ensuring the correctness.
In the next section, we elaborate on ordering linearizability, which is used for developing a specific version of fairness.

\section{Routing Protocols}
\label{sec:route}

At the high level, the goal of a routing protocol is to deterministically assign nodes to hubs, and to create a fixed number of random paths for transactions to traverse.
After a client submits copies of a transaction to expected number of paths, it is the goal of the routing protocol that there is a constant probability that at least one of the paths is made of entirely regular hubs. 
Similarly for adversary, the protocol ensures there is negligible probability that there is a path made of entirely corrupted hubs, even if the adversary sends the transaction to all possible paths.
Those paths are randomly generated for preventing adversary from manipulating the creation process.
In every block, there are only a fixed number of path available regardless of transactions, so that no adversary can grind its transactions to discover the corrupted path.
%
%Otherwise, the adversary can mutate its transaction in infinite ways to discover its preferred path.
%
%Because the protocol neither knows what transactions are malicious that it wants to front-run others, nor which nodes are malicious.
%
To achieve the above property, the protocol design makes use a key assumption from the standard BFT protocol that there are at most $1/3$ malicious nodes.
In the following, we show how to use a common technique in computer science called boosting to design the routing protocols.
Due to the size limitation, we state the properties of the routing protocol in Appendix~\ref{appendix:route:routing_property}. The path traversal mechanism is stated in Appendix~\ref{appendix:route:non_singular:traversal}. The Assignment functions in Appendix~\ref{appendix:route:assignment}.

\subsection{Technique of Boosting}
\label{sec:route:boosting}
Suppose there are two types of hubs: a regular hub which occurs at probability of $p_h$, and a corrupted hub with the probability $p_d$, such that $p_h > p_d$. 
Since $p_h > p_d$, we can represent $p_{d}^{\rho} = p_{h}$, for some $\rho < 1$. By arranging the term, we got $\rho=\frac{\log{1/p_h}}{\log{1/p_d}}$.
Suppose all path has a length of $k$,
the probability of a path of entire honest nodes is therefore $g_h = p_{h}^{k}$, and for a path of corrupted hubs is $g_d = p_{d}^{k}$.
As length of path $k$ increases, the probability of occurrence for both type becomes negligible.
Although both are very small, there is a large gap between the two small number. With some calculation, we can show $g_{d}^{\rho} = g_{h}$ or equivalently $\frac{\log{1/g_h}}{\log{1/g_d}}=\rho$.
Suppose all paths are created independently, as the client can try distinct paths linear number of time, the probability for both type increases linearly.
Up to some point after trying $L$ different paths, there is a constant probability the client would hit a honest path.
But since there is a large gap between the two probabilities $g_h$ and $g_d$.
It is possible to find a $L$, such that the product $g_h L = O(1)$, but $g_d L$ is still negligible.
By keeping the total number of possible path low enough, such that there is no much room beyond $L$ tries, 
we can keep the chance for any adversary to find any corrupted paths negligible.
In the following, we demonstrate two different designs of the routing protocol using this common technique. However, as the hub size and path length can be adjusted, there are much more possible design there.

\subsection{Path of Singleton}
\label{sec:route:singleton}
Suppose every node becomes its own hub, then by BFT adversary assumption, $p_d = 1/3$ and $p_h=2/3$. As the result, $\rho=0.369$.
Because we want $g_d = {p_{d}^k} = \frac{1}{n^\tau}$ be negligible, where ($\tau \ge 1$). We can rearrange the term to get the path length $k=\frac{\tau \log{n}}{\log{1/p_d}}$, then probability of both paths can be computed as 
\begin{align}
    \label{eq:1}
    g_h &= {p_h}^{\frac{\tau \log{n}}{\log{(1/ p_d)}}} 
                      =  n^{- \frac{\log{(1 / p_h)}}{\log{(1 / p_d)}}\tau } 
                      = n^{-\rho \tau } = n^{-0.369\tau} \\
    g_d &=  n^{-\tau}
\end{align}
The second equality in Eq(\ref{eq:1}) comes from the logrithmic rule $a^{\log{b}}= b^{\log{a}}$.
By limiting the total number possible paths to $o(n^{\tau})$, the probability is always negligible for adversary to get a path made of entirely malicious nodes.
If a client retry $L$ different paths, the lower bound for the honest client that at least one path is regular is $1 - (1-g_h)^L$.
By adjusting $L$, we can ensure there is a constant probability which one of the transaction will go through a regular path made of entirely honest nodes.

With those numbers, we can easily build a simple protocol with communication complexity of $O(n^{0.369(1+c)})$ with $\epsilon=1/n^c$.
However, in order to ensure the communication complexity be sublinear, we require $c < 1.71$. 
The system allows $n$ total possible path, such that every node is a starting point of some path.
By letting $\tau=c+1$, although adversary can try $n$ different times, by taking the union bound the upper bound probability for adversary is $n^{-c}$. On the other hand, if a client sends transaction to $n^{0.369(c+1)}$ nodes. The probability with at least one regular path is $1 - (1- g_h)^L$.
Suppose we have $n=200, c=1.2$, then the probability with at least one regular path is $0.576$, with $L=n^{0.8118}$.
To further reduce the complexity, we need to increase the hub size.

\subsection{Non-Singleton hubs}
\label{sec:route:non_singular}
%In the previous section, we show a simple protocol whose hub size contains only one node. It already gives a guideline on how to change from the $O(n)$ to sublinear communication complexity. In this section, we reduce it further by increasing the size of hubs and decreasing the length of path.

In the last section, $p_h$ and $p_d$ only differs by a constant factor, so the approach we use to differentiate them is by increasing the path length in order to increase the gap between the two probability.
However, if $p_h$ and $p_d$ already differs by more than constant factor, we can reduce the length of path and therefore decrease the number of tries for transactions to find a regular path.

\begin{figure}[h]
\centering
\includegraphics[width=1\textwidth]{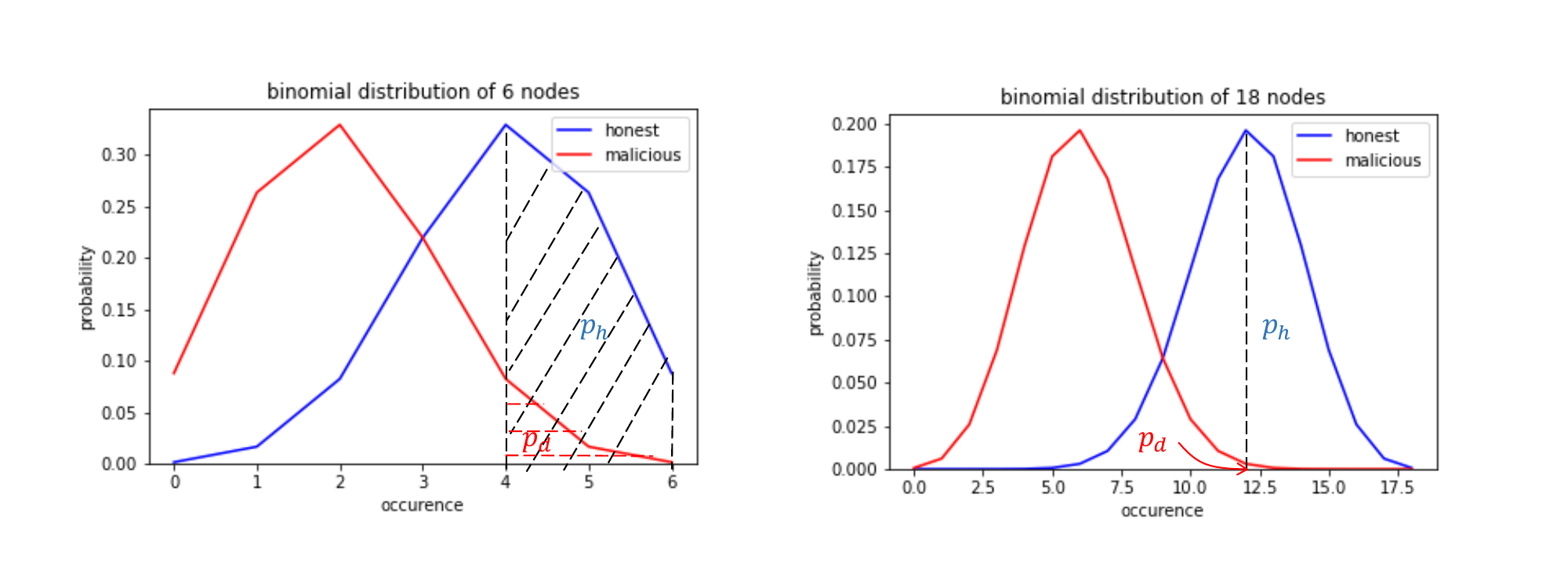}
\caption{Hub Distributions}
\label{fig:hub_dist}
\end{figure}

Suppose now each hub is made of $q$ distinct nodes and every node is draw independently with replacement. 
We know that it is binomial distribution with mean equal to $2/3$, and the binomial distribution is a good approximation to a distribution when nodes are draw without replacement if $n$ is much larger than $q$.
%
%But for simplifying the analysis, for large enough $n$, we can approximate distribution which a hub is made of $x$ honest nodes as a binomial distribution, with $p=2/3$. 
%
Similarly for the distribution for malicious nodes, it can also be modeled as a binomial distribution with mean equal to $1/3$.
The left diagram of Figure~\ref{fig:hub_dist} displays both distributions, when $q=6$. 
Since the hub is not singleton, we need to define a threshold, $t$, which is the minimal number of signatures to get approved by the hub.
A small $t$ makes it easy to pass a hub, but it also increases the chance that a hub gets corrupted. 
We start with $t=\frac{2n}{3}$.
%and it is shown in the left Figure at $t=4$. 
%
%In Figure~\ref{fig:hub_dist}(a) plots the binomial distribution for $n=6$. 
%
%A hub is a regular hub if the number of honest nodes is greater or equal to $t$.
In the left Figure~\ref{fig:hub_dist}, we denotes the mass of probability for passing event in the shaded area.
However, as we increase the size of hub, as shown in Figure~\ref{fig:hub_dist}(b), while keeping $t=\frac{2n}{3}$, the difference between $p_h$ and $p_d$ becomes much more significant.
Since $t$ is equal to the mean of binomial distribution for honest nodes, $p_h= 1/2$.
For computing $p_d$, we can derive it from Chernoff bound (\cite{arratia1989tutorial} Theorem 1), where $D$ is the Kullback-Liebler distance, and $p=1/3$
\begin{align}
    \label{eq:3}
    p_d &\le \exp\{-q D(2/3 || 1/3)\} \\
        &=   \exp \{-q (\frac{2}{3}\log{\frac{2}{3p}}+ \frac{1}{3}\log{\frac{1}{3(1-p)}})\} = \exp\{ -\frac{q}{3} \}
\end{align}

By letting $q=3\log{n}$, we have $p_d = n^{-1}$. To arrive at $\epsilon=1/n^c$, there are two approaches. We can let the path length $k=c$. Or alternatively, we could let $q=3(c+1)\log{n}$, let $k$ be a small constant.
If we choose choose $k=2$ and let $t=2q/3$, then $p_h = 1/2$. The client only needs to submit its transaction to eight different path to ensure one regular path with $90\%$ probability.

%If we let the total possible path be $n$, then the probability of having a malicious path is $n \frac{1}{n^{c}}= n^{c-1}$.
%
In summary, by letting $t=2q/3$ and $k$ be a small constant, we can achieve $O(c\log{n})$ communication complexity to ensure an error probability of $\epsilon=1/n^c$.
However, the protocol does not have to set $t=2q/3$, the threshold can be adjusted by the system to balance between $p_d$ and $p_h$.

%However, if we can either decrease the probability of a corrupted path, we can reduce the length of a path. As the result, it requires less paths samples to have a regular path.
%
%Similarly, we can achieve the same effect, if we can increase the gap between the two types of paths. We might not be able to decrease $h$, but since $h=O(\log{n})$, decreasing it does not have a significantly changes in the communication cost.

%

%The threshold and hub size are two design parameters that trades off liveness and communication complexity.
%
%The threshold determines how easy to approve a hub. 
%
%By setting the threshold low, it is very easy to obtain approval, but it also increases the likelihood of a corrupted hubs. See Figure~\ref{fig:hub_dist}.
%
%A reasonable choice would be setting threshold as $0.5$, as it is the middle part even divides the type 1 and 2 errors. 

%Another design parameter is the hub size. As we increase it, both the type 1 and 2 error decreases, as governed by the binomial distribution. We plot the similar graph below for comparison.

%Given cumulative mass function (cmf) of a binomial distribution, we can calculate the exact probability.

%With a hub size of $\log{n}$, the probability of creating a corrupted hub can be significant less than creating regular hub.     
%
%As the result, the length of a path can be reduced to $O(1)$. A transaction only need a constant number of path in order to have high probability of regular path and negligible probability of corrupted path.

\subsection{Timestamp rendered by types of paths}
\label{sec:route:hubs_timestamps_type}
%In the linearizable fair ordering, the location of a transaction in a block is determined by timestamps of when it is received by nodes. 
%
%For example, {\pompe} uses all $O(n)$ arrival timestamps to derive a medium timestamp which are used to locate itself in the block.
%
The compositions of a path affect the properties of final timestamps. In this section, we delineate all types of hubs, and analyzes how they affect the properties of a path and the resulting timestamps.
Understanding them is important to reason about how {\name} defend against the MEV attacks in the Section~\ref{sec:travelers:strategy}.

\begin{figure}
\centering
\begin{minipage}{.5\textwidth}
  \centering
  \includegraphics[width=1\linewidth]{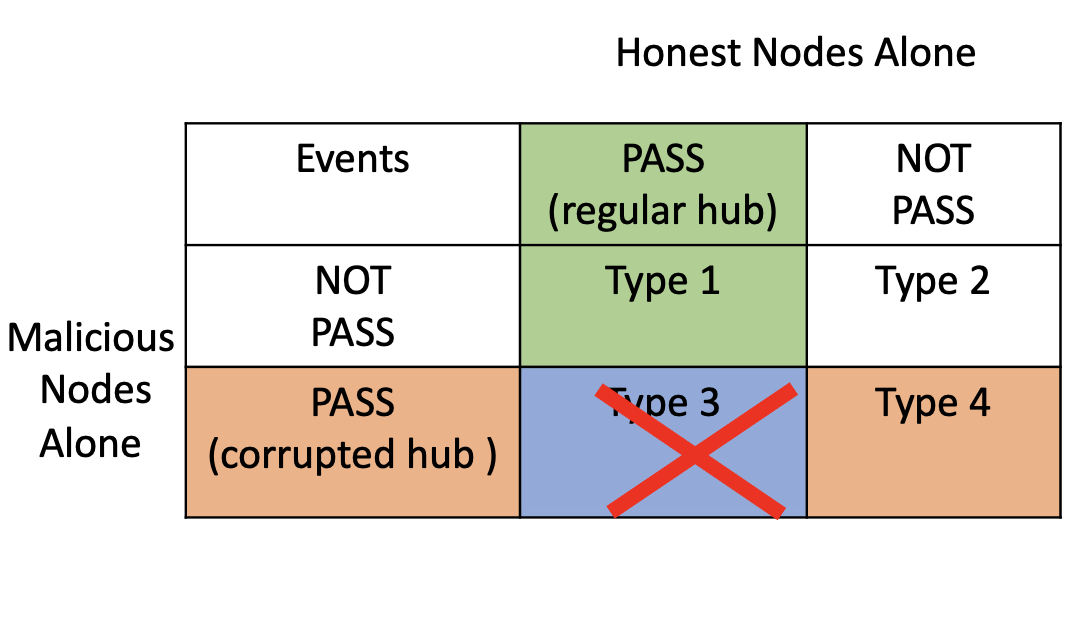}
  \captionof{figure}{Four types of Hubs}
  \label{fig:hub_table}
\end{minipage}%
\begin{minipage}{.5\textwidth}
  \centering
  \includegraphics[width=1\linewidth]{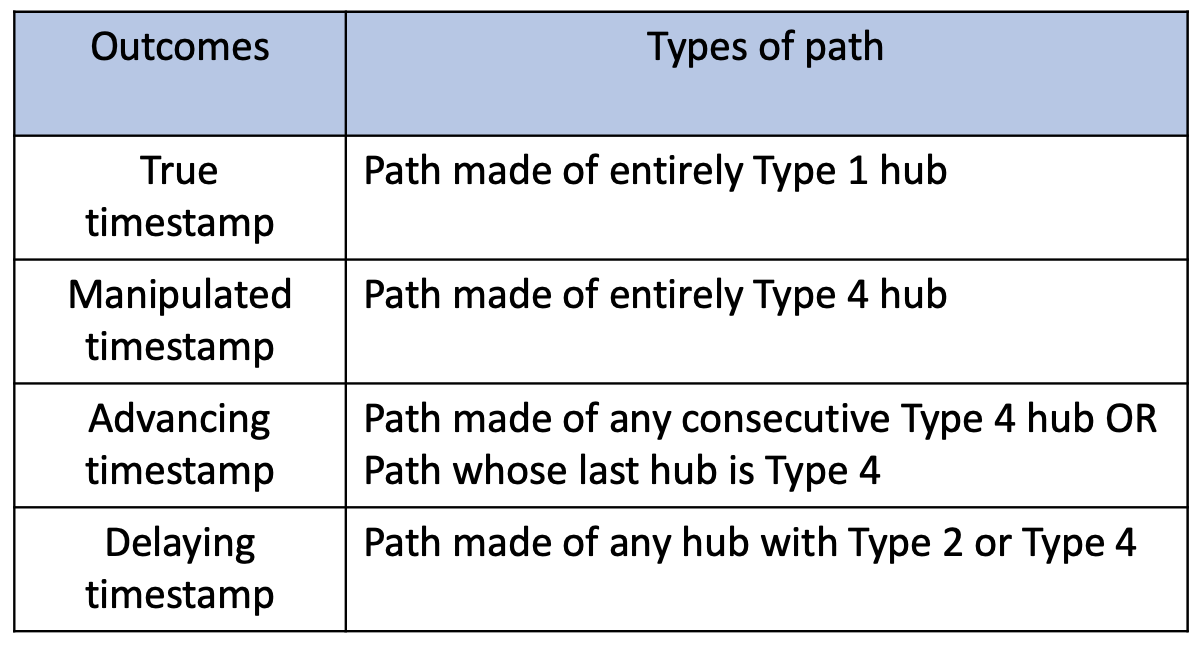}
  \captionof{figure}{Timestamp Types to paths}
  \label{fig:timestamp_table}
\end{minipage}
\end{figure}

%Recall, a path of a transaction is made of a mixture of regular and corrupted hubs. 
%
%Depending on the composition of a path, the generated timestamps would have distinct properties.
%
We start by examining the definition of hubs.
Suppose the size of a hub be $q$ and the passing threshold be $t$. Depending on how many nodes of each type are present in the hub compared to $t$. Let's define an binary event {\textbf{PASS}} by comparing if certain types of node alone can approve the transaction, i.e. if the number of certain type nodes is at least $t$. We can define four types of hubs, as shown in in Figure~\ref{fig:hub_table}.
%

%\begin{figure}[h]
%\centering
%\includegraphics[width=0.9\textwidth]{figs/hub-path-types.png}
%\caption{*}
%\label{fig:hub_path_type}
%\end{figure}

A hub of Type 3 occurs when both honest and malicious nodes alone can approve the transaction. 
But since we want to decrease the chance of a corrupted hub, 
we can simply rmove this case by setting $t>q/2$.
A hub of Type 1 is equivalent to our previous definition of regular hub, and a hub of Type 4 is equivalent to our definition of corrupted hub. 
%
%However, it is undesirable, because if malicious nodes alone can approve the transaction, it increases the chance of a path made of entirely corrupted hubs. In the case, adversary can generate arbitrary timestamp without detection.
%
%Type 4 can be prevented by increasing $t>q/2$. Similarly, the probability of Type 2 hub can be decreased by further increasing $t$.
%
%
%So far, we have casually mentioned the regular hubs as Type 1 hubs, and corrupted hub as Type 4 hubs. By setting $t>q/2$, our previous terminology still holds.
%
The last type of hub we have not explored is Type 2, when a hub is in the situation of impasse if neither honest and malicious nodes alone can approve the transactions, and we want to avoid it for liveness reason.

After defining all types of hub, we can start analyzing types of path made of those hubs. 
In a path made of corrupted hubs,
%and it provides the most advantage to the front-running transaction.
%
a transaction does not need to be sent on the network to get approval,
because the adversary who holds the private keys can approve everything in a central location, any possible timestamp can be manipulated as long as the timestamp does not violate consensus.
%(for example a generated timestamp can only appear in the previously confirmed blocks). 
%
But this is rare by lowering $\epsilon$.
Similarly, a regular path is made of regular hubs. It produces a true timestamp reflective of current network and clock condition. 
But the most common case are paths made of a mixture of corrupted and regular hubs. 

When adversary wants to delay an transaction, as long as there exist one hub along the path that honest nodes cannot approve the transaction alone, the adversary can control to delay the transactions.
A delayed timestamp must have a timestamp later than the earliest true timestamp, otherwise it is advancing timestamps of the transaction.

Sometimes an adversary wants to speedup the delivery and therefore advance the timestamp. 
Suppose the last $x$ hubs on some path are corrupted type 4 hubs, adversary can use a simple method to make it deliver faster, by reusing the last timestamp from honest hubs.
There is another way to advance the timestamp if there is a consecutive corrupted hubs on the path (not necessarily from the end).  
Since all malicious nodes can be controlled by a single adversary, the first corrupted hub can create the approval for all the following corrupted hubs, and therefore help the transaction moves faster than it should have.
Figure~\ref{fig:timestamp_table} summarizes four types of outcomes depending on how a path is constituent of.

Depending on the purpose of adversary, it can selectively
choose either tools to advance or delay some transactions. 
For sandwich attack, adversary can setup a sandwich trap, and then either advance or delay some victim transaction, in order for the victim fall into the trap.
However, by devising the ordering rule properly, we can prevent the adversary from using delaying timestamp entirely.
%
%i.e. with high probability, the final locked timestamp of client's transaction is the true timestamp.
%
The idea is to always use the earliest timestamp as the canonical timestamp, we will elaborate more in Section~\ref{sec:travelers:ordering} and \ref{sec:travelers:strategy}.
Then the only tactics which adversary can use is to advance the timestamp of the victim's transaction.
But it makes the sandwich attack much harder because adversary has to setup the trap earlier than victim's transaction. 
By the property of automated market maker, the adversary incurs a loss if attack is not successful.

\subsection{Encrypting the transaction}
\label{sec:route:encrypt}

However, if we assume adversary has an advanced infrastructure that is magnitude faster than regular Internet backbone, there is a chance that adversary can peek the content of the victim transaction and still have time to setup a sandwich trap. 
In this section, we look at how to incorporate encryption to delay the time when adversary can know about a transaction, therefore making it impossible to setup a trap early enough.

There are multiple ways to implement such encryption scheme. 
At the high level, there must be at least one hub that can decrypt the entire transaction. 
If the hub contains only one node, the decryption is simple.
However, if the hub size has $q$ nodes, there are exponential signature combinations.
Unless $q$ is small enough, otherwise we need a threshold encryption scheme to make it efficient.
Let's define a decryption set $S$ that contains hubs capable of performing decryption.
The hubs can either be group of nodes capable of threshold decryption, or hub whose size is small enough. For example $6$ choosing $5$ contains only 6 possibilities.
For any path that has the ability to perform decryption, it must contain at least one hub from the decryption set $S$.
Ideally the decryption hub is at the end of the path, so that the routing protocol can hide the transaction as long as possible.
If we use iterative path traversal method discussed in Appendix~\ref{appendix:route:non_singular:traversal}, the initiator needs to provide all approval signatures from the beginning of the path to convince the decryption hub that enough time is spent on traversal.
%
%For a client who wants to the best front-running protection, it needs to find a set of paths whose last hubs are part of the decryption sets.
%
A transaction can be encrypted layer by layer if multiple hubs in the decryption set is available. It offers a better protection in case one decryption hub is corrupted.
%
%It is very similar to the onion encryption used by TOR. but their goals are quite different. 
%
%In our scheme, the plaintext will eventually be available on the blockchain readable by everyone. 
%
Our scheme is slightly different from the threshold encryption used by Shutter or anything that uses commit-and-reveal scheme. 
In our scheme, the plaintext will be available after the last hub, so data can be read immediately without further processing. 
%
%We discuss more about the implementation in the next section.
%
%As we discussed before, front-running a transaction requires a sophisticated infrastructure and user. Although the encryption can prevent it from happening, the extra decryption at each layer would impose higher cost to regular nodes from the hubs. 
%
%There are many different types of transaction on blockchain, and not all of them require front-running protection. 
%
%It is transaction owner's choice to decide how many layer of encryption is needed. 

%or trusted party of the client like initiator, which we described in the Path Traversal Mechanism in Appendix~\ref{appendix:route:non_singular:traversal}.

%By making the hubs along the path known to the sender, transaction can be nestedly encrypted in the beginning.
%
%Before sending the transaction, transaction sender needs to repeatedly encrypt the transaction using public keys of hubs along the path.
%
%As transaction moves along the path, the ciphertext are decrypted layer by layer.
%
%Depending on how late the transaction sender wants to reveal its transaction, the sender can decide the number of layer of encryption.

\section{Travelers}
\label{sec:travelers}

{\name} is a consensus protocol that provides probabilistic fair ordering notion we introduced in Section~\ref{sec:fair_ordering}.
In this section, we first provide an overview on the strategy which {\name} use to prevent ordering attacks.
We then show how to design the ordering rule to support such strategy by removing delayed timestamps.
%
%ordering rule, and then we describe the exact properties satisfied by the  {\name} protocol.
%
Because ordering linearizability decouples the ordering from the underlying consensus protocol, the BFT protocol is very modular. 
However, it is important to ensure the BFT protocol has censorship resistance.
%needs to ensure the timestamps generated from the regular path is not censored.
%
We provide an overview on the category of BFT protocols that satisfy the requirements. 
%
%and then we design an censorship resistant consensus protocol with $O(n\lambda)$ complexity.
%At last we integrate with the routing protocol to achieve scalable fair ordering protocol with complexity $O(c\log(n)L + n\lambda)$. 
%

\subsection{Strategy to prevent reordering attacks}
\label{sec:travelers:strategy}

In Section~\ref{sec:route:hubs_timestamps_type}, we show that four types of locked timestamps are generated depending on the paths they are from, as shown in Figure~\ref{fig:timestamp_table}.
%
%We want to understand how the ordering rule helps us remove the effects of delayed timestamp,%a.
%
For an adversary, since the true timestamp is undesirable, and manipulated timestamp is bounded with $O(n^{-c})$ probability,
the adversary only has two tools for arranging its attack: by advancing timestamps and by delaying timestamps.
In the Section~\ref{sec:travelers:ordering}, we show by designing proper ordering rules, we can remove the delayed timestamps from the attack vectors, as long as there is at least one true timestamps being committed by the consensus protocol.
So the only tool which an adversary can affect the ordering is using advanced timestamps by speeding up its or victim's transactions.
However, by using encryption in Section~\ref{sec:route:encrypt}, we can ensure that the clients' transactions have completed most of its traversal on the path.
At last, we note it is up to clients to decide how much front-running protection being offered, a worried client would sends transaction to many paths to ensure the true timestamps from the regular hub is not censored, and the client would use the most secure encryption discussed in Section~\ref{sec:route:encrypt} to make sure the transaction is not front-runnable.

\subsection{$1-\kappa$ Probabilistic ordering linearizability Property}
\label{sec:travelers:properties}

We now formally define the probabilistic properties which overall {\name} has, based on the probabilistic fair ordering notion. The proof is provided in Appendix~\ref{appendix:travelers:proof}.
\begin{itemize}
    \item For every new block in the BFT protocol, there is $1/n^c$ probability that some transaction $tx_d$ provided by a corrupted path can be added into the block at any locations. For the remaining $1-1/n^c$ probability, 
    if there is $1 - \kappa$ probability that the timestamps of both transaction $tx_a, tx_b$ provided by some regular paths are committed, then with $1 - \kappa$ probability the following fairness notion holds:
    if the timestamp of $tx_a$ is {\resolution} duration before the timestamp of $tx_b$, then $tx_a$ will occur before $tx_b$ in the final totally ordered ledger, where $k$ is the number of hubs in a path.
\end{itemize}

We added a new probability $1-\kappa$ to ensure that the protocol covers the case that both transactions are committed.
Im compute the resolution {\resolution}, the protocol uses the iterative routing, which doubles the latency to traverse a path.
If we choose to use a leaderless asynchronous BFT protocol which is inherently censorship resistance, then $\kappa$ can be decreased to negligible. 
DispeseredLedger\cite{yang2022dispersedledger} invents a technique called inter-node linking that guarantees every collected transaction is delivered.   
Similarly, all DAG-based protocols can fetch transactions that were submitted but not confirmed in the past, and confirm them in the new blocks.
However,it requires longer confirmation latency, which we discuss in Section~\ref{sec:travelers:ordering}.
%

%
%Since all leaderless protocol has communication complexity of $O(n^2)$, a leader based BFT protocol is preferred.
%whose censorship resistance depends on how much resources a client wants to spend on the transaction. 

\subsection{Ordering Rule}
\label{sec:travelers:ordering}

When a client sends a transaction on multiple paths, the same transaction would have been locked many timestamps. 
The ordering rule decides which one becomes the canonical timestamp used by the totally ordered ledger.
%
%The core observation is that given there is a true timestamp $t_{true}$ and a resolution of $2(h-1)\Delta + 2\delta$ duration on both before and after sides.
%
%A delayed timestamp is only significant after $t_{true} + 2(h-1)\Delta + 2\delta$.
%

The core observation is that we always have a set of true timestamps committed in the final ledger. By letting the earliest time among them as the canonical timestamps, we filter out all any delayed timestamps. By definition any timestamps that is earlier than the earliest true timestamp belongs to the category of advancing timestamps.
%
%Although it is possible that delayed timestamp might be earlier than the true timestamp if adversary has good Internet infrastructure.
%
%true timestamp at $t_{true}$, any timestamps that wants to delay the delivery must come after $t_{true}$
%
%By choosing the earliest timestamp among all committed timestamp, we can prevent all delayed timestamps from manipulating the final locked timestamp by choosing the early timestamps.
%
%In {\name}, we use the earliest timestamp.
%

%On the other hand, the adversary can use advanced timestamps to prepare its attack, either speedup the victim transactions or the frontrunning transactions.
%
%But by the encryption technique we introduced in Section~\ref{sec:route:encrypt}, 
%
%the transaction can start traversing the the path,
%while the reveal of a transaction can be delayed later.
%
%It is like a racing track, but the revealed transaction needs to travel less distance because those distance has been traveled by the encrypted transactions.

%Ordering linearizability has a unique property compared to batch-order-fairness, which can significantly simplify the consensus protocol.
%
%After completing the routing protocol, multiple timestamps are locked with the transaction stored across the nodes. 
%
Note that the protocol does not need to process anything to 
specify which timestamp is canonical for each transaction.
It is because if a party retrieves all confirmed transactions from the BFT protocol, the party can collects all the timestamps corresponding to a transaction and determine the canonical timestamp by a sorting algorithm.
For this reason, consensus is simply a matter of arriving agreement on a stream of bit, which is much simpler than deducing Condorcet cycle and mitigating the nested loops.

Sometimes there might be a new transaction that locks to a timestamp earlier than what is already confirmed in the past. 
There are two methods to deal with it. The simplest approach is to add the transaction to the latest unconfirmed block with a new timestamp as early as possible without violating the block boundary.
Another way would be altering confirmation rule at the application layer, such that the most recent $x$ block are subject to changes, depending the timestamps of transactions in the new block.
%the current block number minus $x$ are confirmed, but for the recent $x$ blocks, the order is subject to change according the sorted times.
%
However, we have to modify the property a bit to cover the effect when there is a corrupted path.

\subsection{Consensus and Censorship Resistance}
\label{sec:travelers:consensus}

{\name} is intended to be modular systems. 
%So far we have combined the consensus protocol and the fair ordering protocol. 
%
However, there is another desired property called censorship resistance, which is not covered by most BFT protocol.
%
%A censorship resistance protocol guarantees some transactions is included in the next block even if the current and future leaders have the intent of censoring.
%
It is important for {\name}, because if the leader is malicious, it can selectively remove the earlier timestamps, so that only the delay timestamped manipulated by the adversary is regarded the canonical timestamp for the transactions.
Then the adversary can setup a sandwich and delay the victim transactions to fall into the trap.

It can be fixed by requiring the underlying consensus protocol be censorship resistance.
%
%\begin{itemize}
%    \item Censorship resistance: all transactions that are fairly ordered cannot be censored.
%    \item Ordering Consensus: all nodes come to agreement about either absolute reception time or relative ordering among other transactions.
%\end{itemize}
For example, most leaderless protocols are censorship resistant.
%but every node broadcast its information to other nodes, in order to arrive at consensus. 
%
But they all have at least communication complexity of $O(n^2)$ because of all-to-all communication pattern.
Both DispersedLedger and Dumbo-MVBA\cite{mvba} use erasure code to make agreement on a stream of bits without downloading the whole data.
Dumbo-MVBA has a communication complexity of $O(n^2)$.
%We now provide a scheme on how to design a leader based censorship resistance protocol.
%whose censorship resistance depends on how much resources a client wants to spend on the transaction. 
BigDipper provides a more systematic presentation, and offers $O(n^2)$ or $O(n\log{n})$ communication complexity.

\section{Conclusion}
\label{sec:conclusion}
We present {\name} composed of a flexible routing protocol and a censorship resistant BFT protocol.
It achieves a notion called probabilistic ordering linearizability with $O(c\log(n)L + n^2)$ communication communication complexity.
However, the probabilistic notion can also be applied to batch-order-fairness. It will be interesting to discover  new protocols based on relative transaction order.

%Althought this work focuses on the notion of ordering linearizability, it is possible to imagine similar probabilistic attribute can be added to the other group of notion based on batch order fairness. 
%

%Similar probabilistic attribute can be added to the notion of batch order fairness, however, it is not the primary focus of this work.

%
% ---- Bibliography ----
%
% BibTeX users should specify bibliography style 'splncs04'.
% References will then be sorted and formatted in the correct style.
%
\bibliographystyle{splncs04}
\bibliography{references}

\section{Complexity Analysis}
\label{appendix:complexity}
\subsection{Themis}
\label{appendix:complexity:themis}
Themis requires a desemination network, such that everyone can receive the transaction of size $L$, therefore the total communication complexity is $O(nL)$. 
In the consensus part, suppose there are $T$ total transaction, each node has to provide the relative ordering among the $T$ transactions to the leader. So the leader has a complexity of $O(nT\lambda)$, where $\lambda$ is the hash digest to used to denote a transaction among the relative ordering. 
Because the leader in the practical non-SNARK version needs to send the relative ordering of all nodes to every node,
the total complexity is $O(n^2T)$ for consensus, and $O(nTL)$ for transaction submission.

\subsection{Quick Ordering Fairness}
\label{appendix:complexity:quick}
Quick Ordering Fairness protocol consists of a BCCH (Byzantine FIFO consistent broadcast channel) and a VBC (validated byzantine consensus).
To send one transaction, BCCH requires $nL + n^2\lambda$ communication complexity, as provided by the Quick Ordering Fairness paper\cite{cachin2022quick} itself.
The VBC part is modular, the most efficient asynchronous consensus requires $O(n^2)$. Although it can also be used with partial synchronous protocol, its complexity is dominated by the BCCH.

\subsection{{\pompe}}
\label{appendix:complexity:pompe}

The {\pompe} protocol also requires all nodes to receive a transaction.
There is $O(n^2)$ term, because there is a \textbf{Sequence} message to all nodes, and the message contains collected timestamps from $O(n)$ nodes.
It is required to let every node to make agreement on the median timestamp.
For {\pompe}, it uses a standard leader based Hotstuff protocol, which does not offer censorship resistance, and its the complexity is simply $O(n\lambda)$.
Overall the complexity for $T$ transactions is $(n^2TL + n\lambda)$.

\subsection{Travelers}
\label{appendix:complexity:travelers}

In travelers, a transaction passes the the submission protocol if it can traverse a path. 
In Section~\ref{sec:route:non_singular}, we show if the hub size is $O(\log{n})$, and the length of path is constant. A client can send the transaction to constant number of path.
The overall complexity per transaction is $O(\log{n}(\lambda)$.

In Table~\ref{table:ordering-notion}, we list two versions of Travelers.
In Travelers-Speed, the protocol uses recursive routing, so that the same transaction is sent every node in the hub, therefore it is $O(\log(n)(\lambda+L))$.
Recursive routing is useful for reducing the time for the traversal time.
See Appendix~\ref{appendix:route:non_singular:traversal} for more information.
In evaluation the complexity for consensus, we can either use VABA\cite{abraham2018validated} or Dumbo-MVBA\cite{mvba}, they require $O(n^2)$ communication for reaching consensus on a stream of bit.
In BigDipper, it contains a protocol with consensus complexity of $O(n^2)$.

In the other version, Travelers-Light optimizes for low communication complexity. It uses iterative routing, so only the hash of transaction needs to be sent to nodes, which is used for signing.
The initiator in the recursive routing can just collect the signatures, and deliver the transaction along with the signatures to the consensus protocol.
Therefore the communication complexity per transaction is $O(c\log{n}\lambda + L)$.
In BigDipper, we show it is possible to design a censorship resistance protocol with total communication complexity of $O(\log(n)n)$.
Each node only needs to spend $O(x/n + \log{n})$ complexity in consensus, where $x$ is total amount of data required to be reached agreement on after traversing the routing protocol from all nodes, which equals to $\log(n)TL$.
The overall complexity for $T$ transaction is therefore $O((c\log{n}\lambda + L)T + \log(n)n)$.

\section{Routing Protocol}
\subsection{Routing Protocol Properties}
\label{appendix:route:routing_property}

The routing protocol satisfy the following properties:
\begin{itemize}
    \item $O(1)$ Probabilistic Regular path: For any transaction, there is $O(1)$ probability that the protocol can generate a set of timestamps from a path made of entire regular hubs.
    \item $\epsilon$ Probabilistic Corrupted path: In each block, there is only $\epsilon$ probability that an adversary can arbitrarily create certificate with any metadata for all transactions in the block.
%\end{itemize}
\end{itemize}

\subsection{Path Traversal Mechanism}
\label{appendix:route:non_singular:traversal}

%The routing protocol predefines the total number of possible path, so that the adversary cannot grind a corrupted path.
%
%For the rest of paper, let's assume there are $n$ possible total paths, and every node is a start to some path.
%
%Each node uses the deterministic assignment function to compute the all hubs along its path, and all nodes within each hub.
%
In this section, we delineate the procedure how a transaction starts to be processed by the system and how a hub delivers its approval to the next hub.
When a client sends a few copies of its transaction to the routing protocol, with high probability there is at least one honest node following the protocol to forward the transaction to the next hubs. Let's call the head of a path the \textbf{initiator}. The initiator is always a single node.
If the initiator is trusted by the client, it can simplify protocols like encryption in Section~\ref{sec:route:encrypt}.

%
%Suppose the size of hub is $q$ and the threshold be $t$, in most non-trivial case when $q \neq 1$, 
%
Since a threshold of signatures are required for an approval from a hub,
we need an efficient way to aggregate those signatures, because all signatures need to be carried to the end,
which can be verified to show the transaction has traversed the entire path.
We choose to use BLS signature for its property to combine signature. 
Once the first nodes in the first hub receives the transaction, honest nodes in the hub need to combine the  signature and deliver them to nodes in the next  hub. 
%There are two mechanism to achieve the purpose.

\subsubsection{Iterative vs. Recursive routing}
Suppose hub $A$ is a regular hub that has sufficient honest nodes to approve the transaction. 
There are multiple solutions to solve the problem. We can categorize them as either iterative or recursive solution. 
When using iterative queries, like DNS, the initiator sends hashes of transactions to every nodes in the hub, and waits to collect any aggregate signature. The initiator repeats for all hubs.
This approach increases the latency to finish a query and increases the workload for the initiator. But it has benefits of being simple.
%because the initiator can aggregate those signatures and move to the next hub. 
Because the initiator is the single entity, we can easily add more features like encryption we mentioned later.
It also offers a benefits of communication efficient, because to create a signature, the node does not need the entire transaction, but just its hash digest.

%It also has good programmability on the initiator to provide additional features.
%

On the other hand, in a recursive routing, the initiator only participates in the beginning. Once sufficient honest nodes in the first hub receives the transactions, the hub works with the next hub directly to deliver the signature.
The solution requires all honest nodes in the hub to broadcast its signatures to all nodes in the next hubs,
once sufficient honest nodes in the next hub collect sufficient signatures, they combine the signature from previous hub locally and repeat the same procedures until the end of a path.
This approach is faster in latency but requires $q^2$ communication complexity per hub.
All signatures along the hub needs to be carried to the last hub, so that the consensus is able to verify that the transaction has traversed sufficient hubs.

\subsection{Assignment functions}
\label{appendix:route:assignment}

So far, we simply assume there is a hub assignment function from nodes to hub, and a path assignment function from hubs to path.
In this section, we provide a general description on those functions.
Let $\mathcal{D}$ denotes the set of all nodes, $\mathcal{H}$ denotes be the power set (set of all possible subset) of $\mathcal{D}$.
Let's denote the hub assignment functions as $f: \mathcal{D} \rightarrow {\mathcal{H}}$.
Let $k$ be the path length for all paths, then the path assignment function is $g: \mathcal{H} \rightarrow \mathcal{H}^k$.

Multiple ways are possible to create those assignment function. One method requires the use of a random source that produces a random number in every block, let's denote the randomness as $r$. 
For block number $b$, node $i$ can computes $q$ different members in the $j$-th hub by taking $Hash(i, j, k, b, r) \; mod\; n$, where $n$ is number of nodes, and $0<k<q$ (if one node is selected twice, then replace $k$ with $2k,3k..$ until a distinct node is found).
Then repeat the procedure for $0<j<k$, to obtain all nodes on its path.
Because the randomness is globally visible by every nodes, all nodes would be able to obtain all paths by computing locally.
Since eac
Randao in Ethereum currently offer such randomness.

Another method does not require a global visible randomness, it can be achieved with verifiable random function (VRF)). A VRF is cryptographic tool that takes inputs of a seed and secret key of a key pair, and outputs a random number and a proof that the number is generated correctly.
Every node then uses its private key as the source of randomness and generate all the nodes on its path. 

%However, using hash function to create the assignment function is only an example. In principle, one can create hub of different size, as long as the probability of creating a corrupted path is low.
%

Because the node assignment function can be very generic, we can add some structure to the function for enabling some features. 
For example, if we know some group of nodes already have the threshold signature setup, and the group of nodes has sufficiently high  security assumption.
We can group those nodes as a super node, so that when creating a new hub, if the node assignment function pick the super node, it would not select other nodes that is outside the group to be part of the new hub.
In Obol\cite{obol}, it is possible to have a threshold setup with 7 honest out of 10 nodes.

We can also change the path assignment functions to improve encryption scheme from Section~\ref{sec:route:encrypt}.
For example, the assignment function always include a hub from the decryption hub set in the end of every path.
The assignment function can also change the size of hubs on a path, as long as the final probabilities for both corrupted path and regular path satisfy the protocol requirement.
Some smaller hubs are therefore eligible to be a part of decryption hub set.

\section{Travelers Consensus}
\label{appendix:travelers}

\subsection{Property Proof}
\label{appendix:travelers:proof}
An adversary can arbitrarily add new transaction into any location in the ledger if it can find a corrupted paths. 
Otherwise it will only be able to advance or delay the timestamps.
By the properties of the routing protocol, there is the error probability of $\epsilon=1/n^c$ for some $c\ge 1$ for that to happen.
In the rest of case, no such event can happen.

Suppose there is $1-\kappa$ probability for a transaction $tx_a$ to get included in the ledger, then by the routing protocol there is $O(1)$ probability that this transaction has a regular path.
Similarly for the other transaction $tx_b$.
Suppose every path has a length of $k$ hubs and network latency is bounded by $\Delta$.
In reference to the true timestamp produced by the regular path, due to the network uncertainty, those two transactions are separable as long as the true timestamps between the two differ by $2k\Delta$.
Suppose we choose the iterative routing mechanism to move from hubs to hubs which is the longer compared to the recursive routing, the total time spent on the traversal is doubled as $4k\Delta$, because of the round trip.

Suppose the mismatch among nodes' clock is bounded by $\delta$, then the maximal mismatch between two true timestamps are $2\delta$.
Hence if transaction $tx_a$ is committed and separated by duration $4k\Delta + 2\delta$ from the transaction $tx_b$, then $tx_a$ is ordered before $tx_b$.

\end{document}